\documentclass[%
 reprint,
superscriptaddress,
amsmath,amssymb,
aps,
pra,
floatfix,
]{revtex4-2}

\usepackage[english]{babel}

\usepackage{amsmath,amssymb,amsthm}
\usepackage{tikz}

\usepackage{pgfplots}
\pgfplotsset{width=10cm,compat=1.9}

\usepackage{physics}
\usepackage{graphicx}
\usepackage[colorlinks=true, allcolors=blue]{hyperref}

\definecolor{blau}{RGB}{50,50,150}

\begin{document}
\preprint{APS/123-QED}

\title{Quantum simulation of the Heisenberg XXZ model on a Rydberg atom array}

\author{Matthias Werner}
\email{matthias.werner@qilimanjaro.tech}
\affiliation{Qilimanjaro Quantum Tech., Carrer de Veneçuela, 74, Sant Martí, 08019 Barcelona, Spain}
\affiliation{Departament de Física Quàntica i Astrofísica (FQA), Universitat de Barcelona (UB), Carrer de Martí i Franqués, 1, 08028 Barcelona, Spain}
\affiliation{Institut de Ci\`encies del Cosmos, Universitat de Barcelona, ICCUB, Carrer de Martí i Franqu\`es, 1, 08028 Barcelona, Spain}

\author{Artur García-Sáez}
\affiliation{Qilimanjaro Quantum Tech., Carrer de Veneçuela, 74, Sant Martí, 08019 Barcelona, Spain}
\affiliation{Barcelona Supercomputing Center, Plaça Eusebi Güell, 1-3, 08034 Barcelona, Spain}

\thispagestyle{empty}
\pagestyle{empty}

\begin{abstract}
    A major application of analog quantum hardware is quantum simulation, where a system evolves over time according to a given Hamiltonian. One challenge when implementing the Hamiltonian on a programmable quantum simulator is the availability of distinct coupling types, many of which go beyond the conventional ZZ-couplings found in Ising Hamiltonians that are naturally implemented by various platforms. In prior work, we have developed a method based on domain wall encoding to simulate one-dimensional spin chains using only Ising Hamiltonians, which would otherwise require XX+YY-couplings as well. Here, we implement the method on a commercially accessible Rydberg atom quantum device, and simulate the time evolution of the Heisenberg XXZ model with various anisotropies. We successfully probe the XY and Ising phases of the model and reproduce the qualitative behavior of several one- and two-body observables in both phases, as well as at the critical point. Our work augments the toolbox to simulate canonical quantum many-body systems on analog quantum simulators.

\end{abstract}

\maketitle

\section{Introduction}
Quantum simulation is one of the most anticipated applications of programmable quantum hardware. Various methods and devices have been conceived to perform quantum simulation of complex many-body systems~\cite{Abrams_1997, Ortiz_2001, Bravyi_2002}. Over time, the available hardware has become increasingly coherent, making it possible to demonstrate quantum mechanical behaviors in a controlled manner \cite{Bernien_2017, Harris_2018, Xu_2020, King_2018, King_2022, King_2023, Li_2023}.

Quantum simulators based on Rydberg atom arrays have become a promising platform for analog quantum simulation. While many analog quantum simulators can theoretically be considered universal quantum computers~\cite{Werner_2026, Hu_2026}, in practice simulating spin chains with large anisotropy is challenging on some of these devices. Rydberg atoms give rise to effective Hamiltonians where interactions are often not freely programmable, as the interaction strength and topology are governed by the distance between atoms, and the interaction type depends on the choice of Rydberg states~\cite{Whitlock_2017, Signoles_2021}. Recent theoretical proposals suggest that programmable XXZ interactions could be realized by an appropriate choice of Rydberg states in combination with a magnetic field~\cite{Kunimi_2025, Kunimi_2026}, while experimental approaches typically rely on Floquet engineering~\cite{Scholl_2022} or effective Hamiltonians in subspaces of a particular particle number \cite{Kim_2024}. However, in commercially accessible devices, one is often limited to the Ising type of interactions together with a transverse field.

Quantum spin chains are canonical models that display various intriguing behaviors characteristic of quantum many-body systems, serving as a benchmark system \cite{Chancellor_2022}, or to investigate many-body localization \cite{Barisic_2010, Vardhan_2017, Mierzejewski_2018, Lin_2018, Wang_2024}, topological phases \cite{Ganeshan_2013}, and quantum transport \cite{Sirker_2020, Kandel_2021, Mellak_2023, Chen_2024, Michel_2026}. In a recent paper, we proposed a method to approximately simulate Heisenberg XXZ spin chains to a controlled error using one-dimensional transverse-field Ising chains with nearest-neighbor and next-nearest-neighbor ZZ-interactions \cite{Werner_2024}. This encoding method allows us to overcome the limitations mentioned above and simulate XXZ spin chains with programmable anisotropy using Rydberg atom arrays that realize transverse-field Ising Hamiltonians.

To our knowledge, we demonstrate the first implementation of our method on a Rydberg atom quantum simulator. We simulate the time evolution of the homogeneous open Heisenberg XXZ chain with local boundary fields. This model shows distinct phases with different dynamical properties, which we probe by simulating the time evolution of spin excitations moving along the chain. While analytically solvable, this model can serve as a benchmark problem for quantum simulators.

Our proposed method is capable of reproducing the expected dynamical behavior of the Heisenberg XXZ model and its dependence on the anisotropy. Prior works on this model using low-order Floquet engineering of the native XY interactions are algebraically bounded to a finite range of anisotropies~\cite{Scholl_2022}. Our proposal does not have such a hard ceiling and qualitatively probes anisotropies beyond this bound. Kim et al.~\cite{Kim_2024} realize highly anisotropic spin interactions. However, the realized effective models are equal to the pure Heisenberg XXZ only in the symmetry sector of a single excitation. As we discuss in the following, our approach is able to simulate the Heisenberg XXZ model also for a larger variety of excitation numbers.

The paper is structured as follows. First, we introduce the Hamiltonian of the Rydberg atom quantum simulators. Then we argue that, for an appropriate choice of parameters, it permits an interpretation as an anisotropic XXZ chain using domain wall encoding, and we briefly discuss the properties of the anisotropic XXZ chain that we simulate, as well as modeled error sources and the initialization of the device. This is followed by a brief numerical investigation of the error contributions to the simulated model. Finally, we present the results obtained from the Rydberg atom simulator, after which we conclude and give an outlook on potential improvements and future research.
\begin{figure*}
    \centering
    \includegraphics[scale=.25]{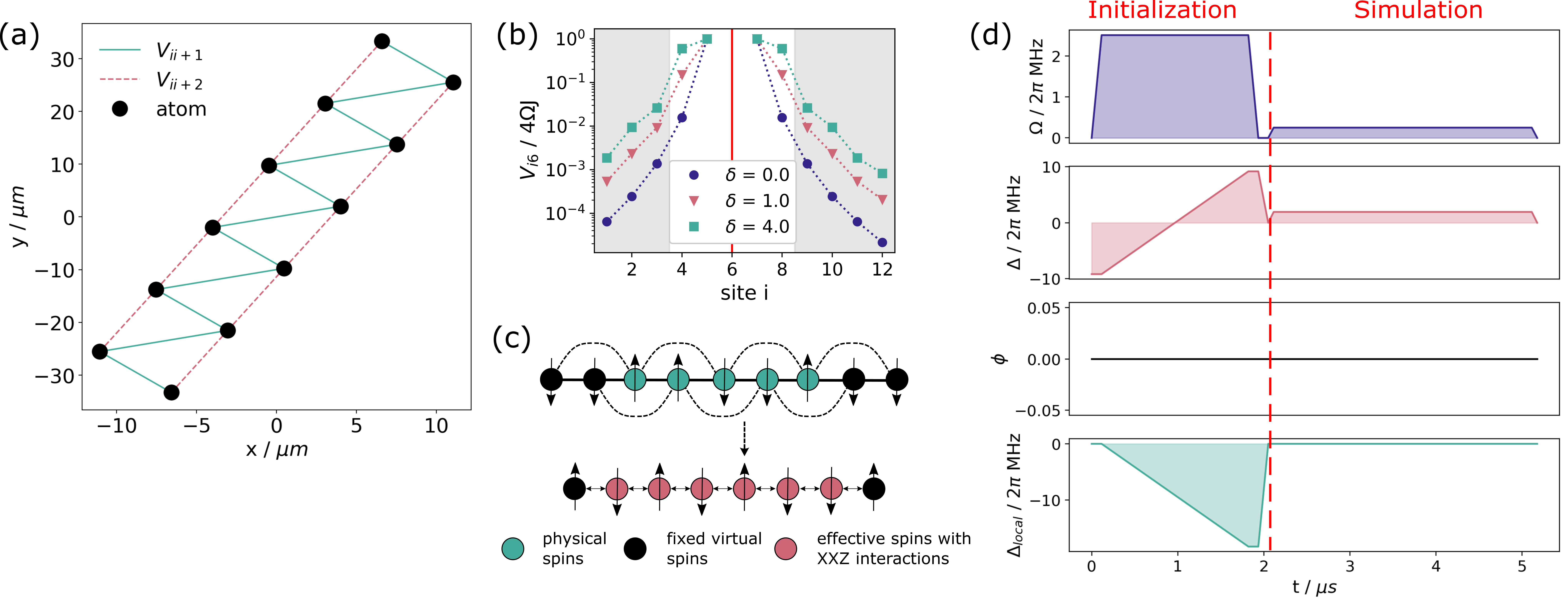}
    \caption{(a): Arrangement of Rydberg atoms in the xy-plane as a chain with nearest-neighbor and next-nearest-neighbor interactions, depicted as solid green and dashed red lines respectively. Other interactions can be neglected due to the decay with the sixth power of the atom distance; (b):  Interaction strength of atom $j=5$ with all other atoms for atom arrangements corresponding to anisotropies $\delta = 0$ (blue dots), $\delta=1.0$ (red triangles) and $\delta=4.0$ (green squares). The white background indicates the nearest neighbor and next-nearest-neighbor interactions, while the gray background indicates all interactions beyond these two. The gray shaded interactions are undesired for all values of $\delta$, except that in the case of $\delta=0.0$ the next-nearest-neighbor interactions should ideally also be zero. However, all the undesired interactions are approximately an order of magnitude smaller than the desired interactions, which justifies neglecting them in the approximate Hamiltonian Eq.~\eqref{eq:RydNNNApprox}; (c) Illustration of the map from an Ising chain with nearest- and next-nearest-neighbor interactions to a quantum spin chain via domain wall encoding; (d): pulse sequence of the driving laser, showing the amplitude $\Omega$ (blue), the global detuning $\Delta$ (red), the phase $\phi$ (black) and the local detuning $\Delta_{local}$ (green) corresponding to the Hamiltonian Eq. \eqref{eq:HRyd} as functions of time.}
    \label{fig:register}
\end{figure*}

\section{Method}
In this section, we first introduce Rydberg atoms as a quantum simulation platform, specifically for simulating the Heisenberg XXZ model with local boundary fields. Since the domain wall encoding for the simulation of spin chains is formulated for Pauli-Z operators $\sigma^z$, instead of Rydberg state occupation number $\hat{n}$, we first show how a particular choice of parameters yields the appropriate Hamiltonian formulated with $\sigma^z$. In a second step, we argue that this Hamiltonian permits a Heisenberg spin chain interpretation.

\subsection{Quantum simulation with Rydberg atoms}
\label{sec:QSimRydberg}
We implement the simulation on QuEra's Aquila device \cite{Wurtz_2023}. The qubits are realized by Rb-87 atoms placed in a two-dimensional plane and subjected to a global Rabi drive with a time dependent amplitude $\Omega(t)$, time dependent phase $\phi(t)$ and time dependent detuning $\Delta(t)$. The interactions $V_{ij}$ are realized by van der Waals interactions and depend on the positions of the atoms $\vec{r}_i$
\begin{equation}
    V_{ij} = \frac{C_6}{|\vec{r}_i - \vec{r}_j|^6}
    \label{eq:RydbergInteractions}
\end{equation}
with $C_6 = 862690 \cdot 2\pi$ MHz $\mu m^6$. The $V_{ij}$ are programmed via the atom position $\vec{r}_i$, which can be set once before each quantum simulation run and are then considered fixed for the remainder of the execution of the analog quantum program.

On Aquila it is also possible to introduce local longitudinal fields, where a pattern $0 \leq h_i \leq 1$ is fixed beforehand, while the time dependence of the local detuning can be collectively controlled on all atoms by the intensity of the local addressing $\Delta_{local}(t)$. In total, the Hamiltonian reads
\begin{equation}
    \begin{aligned}
        &H_{Ryd} = \sum_i \frac{\Omega(t)}{2} \left( e^{i\phi(t)} |0_i\rangle \langle 1_i| + H.c. \right)  \\
        &- \Delta(t) \sum_i \hat{n}_i + \sum_{j>i} V_{ij} \hat{n}_i \hat{n}_j -\Delta_{local}(t) \sum_i h_i \hat{n}_i \ .
    \end{aligned}
    \label{eq:HRyd}
\end{equation}
We will first discuss the static Hamiltonian used for the quantum simulation. Then, we will discuss the initialization procedure, for which we use the local detuning. Since the pattern $h_i$ for the local detuning cannot be changed throughout the simulation, we restrict its use to the preparation of the initial state, whereas, during the simulation, we set $\Delta_{local} = 0$. Furthermore, for the duration of the quantum simulation, we keep the Rabi frequency and the global detuning constant, i.e. $\Omega(t) = \Omega$ and $\Delta(t) = \Delta$. The Rabi frequency $\Omega$ will be the energy scale of the hopping terms in the XXZ chain, therefore, we will use $\Omega$ as the scale for the other interaction energies. We will also keep the phase $\phi(t)=0$ at all times, so that the off-diagonal terms are given by $\sum_i \sigma_i^x$.

The Hamiltonian of the Heisenberg XXZ chain reads
\begin{equation}
    H_{XXZ} \propto \sum_i \sigma_i^x \sigma_{i+1}^x + \sigma_i^y \sigma_{i+1}^y + \delta \sigma_i^z \sigma_{i+1}^z \ .
\end{equation}
In the following, we show how this Hamiltonian can be simulated using domain wall encoding on a Rydberg atom simulator, up to local fields acting on the boundaries.

The Hamiltonian $H_{Ryd}$ can easily be converted to an Ising Hamiltonian by plugging the identity $\hat{n}_n = (1+\sigma^z_n) / 2$ into Eq.\eqref{eq:HRyd}. This results in
\begin{equation}
    \begin{aligned}
        &H_{Ryd} = \frac{\Omega}{2} \sum_i \sigma_i^x - \frac{\Delta}{2} \sum_i \sigma^z_i \\
        &+ \sum_{j>i} \frac{V_{ij}}{4} \left(\sigma_i^z \sigma_j^z + \sigma_i^z + \sigma_j^z  \right) + const .
    \end{aligned}
\end{equation}
The interactions $V_{ij}$ from Eq.~\eqref{eq:RydbergInteractions} are long-ranged and strictly speaking all atoms interact with each other. However, due to the rapid decay of the interaction strength with the atom distance $|\vec{r}_i - \vec{r}_j|^{-6}$, it is possible to obtain good approximations of certain interaction topologies, such as one-dimensional chains. Here we will limit ourselves to approximating an Ising chain with homogeneous nearest-neighbor interactions $\Omega J > 0$ and next-nearest-neighbor interactions $\Omega \delta \geq 0$. The parameter $\delta$ corresponds to the anisotropy of the XXZ chain. We will use it to probe different phases of the chain, as discussed below.

Typically, one implements one-dimensional Ising chains with nearest-neighbor interactions by placing the atoms on a line, where the distance is chosen such that $V_{ii+1} = 4J_{ii+1} = 4\Omega J$. However, in order to realize the ZZ-interactions of the domain walls, we also require next-nearest-neighbor interactions, as we discuss in more detail below. These are realized by arranging the atoms $i$, $i+1$, and $i+2$ in a triangular shape, rather than linearly. The angle of the triangle at atom $i+1$ allows us to control the interaction $V_{ii+2}$. We chose the next-nearest-neighbor interactions to be $V_{ii+2} = \Omega \delta$. The atoms are arranged in a zig-zag shape, such that atoms $i$, $i+1$, and $i+2$ form the appropriate triangle for all $i$. An example of the arrangement is shown in Figure \ref{fig:register}(a). A similar arrangement has been reported in Ref.~\cite{Kunimi_2025} to realize the $J_1$-$J_2$ Heisenberg model.

In this arrangement, the interactions between atoms farther apart than next-nearest neighbor remain negligible, as is shown in Figure~\ref{fig:register}(b), where the unwanted interactions are approximately an order of magnitude smaller than the desired interactions for all values of the anisotropy $\delta$ we consider here. This allows us to drop the interaction terms beyond next-nearest neighbor and consider the approximation
\begin{equation}
    \begin{aligned}
        &H_{Ryd} \approx \frac{\Omega}{2} \sum_i \sigma_i^x - \frac{\Delta}{2} \sum_{i=1}^{N-1} \sigma^z_i  \\ &+ \Omega J \sum_{i=1}^{N-2} \left( \sigma_i^z \sigma_{i+1}^z  + \sigma_i^z + \sigma_{i+1}^z \right) \\
        &+ \frac{\Omega \delta}{4} \sum_{i=1}^{N-3} \left( \sigma_i^z \sigma_{i+2}^z + \sigma_i^z + \sigma_{i+2}^z \right) \ ,
    \end{aligned}
    \label{eq:RydNNNApprox}
\end{equation}
where we have also dropped the inconsequential constants. When transforming $H_{Ryd}$ to an Ising Hamiltonian by change of variables
\begin{equation}
    \hat{n}_i = (1+\sigma_i^z) / 2 \ ,
\end{equation}
the transformation introduces unwanted single-body Z-terms, which require a compensating global detuning $\Delta = \Omega(\delta + 4J)$. This cancels all unwanted fields, except two $J$-fields acting on the edge qubits and four $\delta$-fields acting on qubits $i=1, 2, N-2, N-1$. These remaining fields permit a convenient interpretation in the domain-wall picture, which we will introduce below.

These considerations result in the final Hamiltonian
\begin{equation}
    \begin{aligned}
        &H_{Ryd} \\
        \approx& \Omega \bigg( \frac{1}{2} \sum_i \sigma_i^x - (J + \delta / 4)  \sigma^z_1 - (J + \delta / 4) \sigma^z_{N-1}  \\
        - &\frac{\delta}{4} \sigma_2^z - \frac{\delta}{4} \sigma_{N-2}^z + J\sum_{i} \sigma_i^z \sigma_{i+1}^z + \frac{\delta}{4} \sum_{i} \sigma_i^z \sigma_{i+2}^z \bigg)
    \end{aligned}
    \label{eq:RydFinal}
\end{equation}
The interaction graph of the Hamiltonian is shown in Figure~\ref{fig:register}(c), where the remaining local fields on the end qubits are depicted as couplings to fixed virtual qubits.

\subsection{Domain wall encoding}
We employ the method introduced in Ref.~\cite{Werner_2024}. It is possible to use domain wall encoding \cite{Chancellor_2019} to simulate quantum spin chains with effective hopping terms and longitudinal interactions using an Ising chain with nearest-neighbor and next-nearest-neighbor ZZ-interactions, as we discuss now. The idea of mapping next-nearest-neighbor ZZ-interactions to interactions between domain walls has been used previously to analytically derive effective Hamiltonians of the anisotropic next-nearest-neighbor Ising (ANNNI) model with a transverse field~\cite{Rieger_1996} close to one of its critical points. However, in Ref.~\cite{Werner_2024}, domain wall encoding is used to simulate a large class of versatile spin chain models.

In the formulation used in Ref.~\cite{Werner_2024}, an excitation of the spin chain was encoded in the wall between domains of all qubits in $\ket{0}$ and in $\ket{1}$. The contiguity of the domains is enforced by strong ferromagnetic couplings. However, here we will use strong anti-ferromagnetic couplings, as these are native to Rydberg atom simulators, i.e. the Hamiltonian reads
\begin{equation}
    H_{DW} = -\Omega J \sigma^z_1 -  \Omega J \sigma^z_{N-1} + \Omega J \sum_{i=1}^{N-2} \sigma^z_i \sigma^z_{i+1} \ ,
    \label{eq:HJ}
\end{equation}
where $J >0$. This is a minor deviation from the encoding presented in~\cite{Werner_2024}, but as a consequence, we encode the information in the location of an interface of two $\mathbb{Z}_2$-ordered domains. For example, we can interpret the state $|01011010\rangle \equiv \ket{0001000}_{DW}$ as an excitation at site 4, counting only physical qubits for now. An example is illustrated in Figure \ref{fig:register}(c), where we show also virtual qubits, which we will introduce in the following. More generally, we consider an excitation to be two adjacent qubits in the same state, i.e. $\ket{00} , \ket{11} \equiv \ket{1}_{DW}$. The local Z-fields in Eq.~\eqref{eq:HJ} correspond to couplings to fixed virtual qubits at the ends of the chain. Together with the anti-ferromagnetic ZZ-terms, this enforces that changes in the number of domain walls in the chain, e.g. under dynamics of an added off-diagonal Hamiltonian, are energetically penalized by $\pm 4\Omega J$, so in the limit of $J \rightarrow \infty$, the domain wall number becomes a constant of motion. Note that the domain walls, loosely speaking, live between two qubits. Therefore, considering also the virtual qubits, we require $N-1$ physical qubits for a chain that has $N$ locations to host a domain wall.

Here, we want to simulate Heisenberg spin chains with longitudinal interactions. Therefore, we need to augment $H_{DW}$. Two domain walls can be made to interact by additionally choosing next-nearest-neighbor interactions $\Omega \delta / 4$ in the Ising chain. Consider the three qubit state $\ket{0z0}$, where $z\in\{ 0, 1\}$. Whatever the state of the middle qubit is, it is clear that there is either no domain wall, or there are two. Similarly, if $\ket{0z1}$, it is clear that there is one domain wall in any case. The same argument applies in the cases of $\ket{1z1}$ and $\ket{1z0}$. Therefore, a next-nearest-neighbor ZZ-interaction in the Ising chain is equivalent to a nearest-neighbor ZZ-interaction in the domain wall picture
\begin{equation}
    \sigma^z_i \sigma^z_{i+2} \equiv \sigma_{DW,i}^z \sigma^z_{DW, i+1} \ ,
\end{equation}
and the $\delta$-dependent terms in $H_{Ryd}$ can be interpreted as ZZ-interactions of domain walls.

The transverse fields effectively act as hopping terms on the domain walls. As we have discussed above, for sufficiently large $J$, spin flips that create or destroy domain walls are energetically suppressed. However, a flip of a qubit adjacent to the domain interface moves the excitation to the left or right. As a consequence, $H_{Ryd}$ effectively acts on domain walls as
\begin{equation}
    \begin{aligned}
        &H_{XXZ} \\
        = &\frac{\Omega}{2} \sum_{i=1}^{N-1} (\sigma_{i+1}^+ \sigma_{i}^- + \sigma_{i+1}^- \sigma_{i}^+) + \frac{\Omega \delta}{4}  \sum_{i=1}^{N-1} \sigma_{i}^z \sigma_{i+1}^z \\
        &+ \frac{\Omega \delta}{4} \sigma_1^z + \frac{\Omega \delta}{4} \sigma_N^z \\
        = &\frac{\Omega}{4} \sum_{i=1}^{N-1} (\sigma_{i+1}^x \sigma_{i}^x + \sigma_{i+1}^y \sigma_{i}^y + \delta \sigma_{i}^z \sigma_{i+1}^z ) \\
        &+ \frac{\Omega \delta}{4} \sigma_1^z + \frac{\Omega \delta}{4} \sigma_N^z \\
    \end{aligned}
    \label{eq:HSpinChain}
\end{equation}
up to a constant energy shift within each subspace of constant excitation (domain wall) number~\cite{Werner_2024}. Note that in Eq.~\eqref{eq:HSpinChain}, we drop the subscript ``DW'' on the operators for clarity. Similarly to the local Z-fields on qubits $1$ and $N-1$, the local fields on qubits $2$ and $N-2$ in Eq. \eqref{eq:RydFinal} correspond to couplings to the same virtual fixed qubits, so that the virtual qubits are also coupled to their respective next-nearest neighbor. The Hamiltonian in Eq.~\eqref{eq:HSpinChain} is the Heisenberg XXZ chain with anisotropy~$\delta$.

Note that the virtual fixed qubits on the two boundaries in $H_{Ryd}$ enforce a constraint on the domain wall number that can be realized. Since the virtual qubits in $H_{Ryd}$ are in the ground state $\ket{g}$, the sum of the domain wall number and $N$ needs to be even. Here, we simulate three excitations on $N=13$ sites. While this limits the method to the simulation of an excitation number of a certain parity, the spin chains that can be modeled by the method in Ref.~\cite{Werner_2024} preserve excitation number. Therefore, the respective Hamiltonians are block-diagonal and sectors of distinct excitation number do not interact in the first place. Hence, many-body properties over a large range of excitation numbers can still be probed.

\subsection{The role of anisotropy in the XXZ chain}
Depending on the value of $\delta$, the XXZ model is in the XY phase ($\delta < 1$) or in the Ising phase ($\delta > 1$). The phases are distinguishable by the dynamical behavior of excitations on adjacent sites. In the XY phase, the excitations mostly do not form bound states and propagate ballistically, while in the Ising phase neighboring excitations form bound states and stop moving independently.\\
We will probe the XXZ chain for three values of $\delta$. To simulate the dynamics in the XY phase, we set $\delta = 0.0$, while for the Ising phase we set $\delta = 4.0$. This simulates the anisotropy at twice the limit of previously reported Floquet techniques~\cite{Scholl_2022}. Additionally, we probe the isotropic point at $\delta = 1.0$.

\subsection{Initial state preparation and simulation}
We prepare the initial classical state using local detuning. To this end, we set the Rabi frequency $\Omega = A$ and sweep the global detuning $\Delta$ from $-B/2$ to $B/2$, while at the same time sweeping the local detuning $\Delta_{local}$ from $0$ to $-B$ in time $t_{init}$. The initialization pulse is shown in Figure~\ref{fig:register}(d). The pattern of the local detuning is chosen as
\begin{equation}
	h = (1,1,0,1,0,1,0,1,0,1,0,0) \ ,
    \label{eq:InitialPattern}
\end{equation}
which together with the depicted pulse sequence results in a protocol that approximately adiabatically transforms the system's initial state $\ket{g}^{\otimes N-1}$ into $\ket{\psi_0} = \ket{ggrgrgrgrgrr}$. Ultimately, the Rabi drive, global and local detuning are turned off again ($\Omega = \Delta_{local} = 0$). This concludes the initialization procedure.

For the quantum simulation, we turn the Rabi frequency to the desired hopping amplitude $\Omega$, while the global detuning is set to $\Delta = \Omega (4J + \delta)$, as discussed above. The system is then allowed to evolve for time $t = 0 ... 3\mu s$, after which it is measured in the computational basis. The full pulse sequence is shown in Figure \ref{fig:register}(d).

Due to the sign of the uncompensated local fields on the edge qubits, the virtual qubits are all in the ground state $\ket{g}$, so that the domain wall state prepared by the initialization procedure corresponds to the logical state $\ket{1100000000010}_{DW}$. We chose this state because it allows us to compare the dynamics of an individual excitation, which should move freely for all $\delta$ up to scattering off the boundaries or other excitations, and a pair of excitations, which will form a bound state in the Ising phase. However, in principle, the procedure allows one to initialize any other classical state with the correct domain wall number parity by an appropriate choice of the pattern $h$ in Eq.~\eqref{eq:InitialPattern}.

Since the initial state preparation protocol is based on adiabatic transformation, the finite time execution introduces some errors. We optimize the parameters $A$ and $B$ of the initialization pulse to prepare the initial state with 0.99 fidelity in the shortest possible time by running a unitary simulation of the Hamiltonian Eq.~\eqref{eq:HRyd}. This does not take into account various imperfections of the device, as we discuss in the following section. However, our results suggest that the state preparation procedure works sufficiently well.

\subsection{Error sources of the quantum simulator}
The results obtained from the quantum simulator do not completely match the theoretical predictions by the Heisenberg XXZ chain. There are various imperfections to a real device that can explain the deviations. Here we consider an error model that includes the thermal motion of the atoms, fluctuations in the laser parameters, as well as misclassification errors of the read-out (RO) procedure. The parameters for this error model are taken from calibration data~\cite{Wurtz_2023}. We model the thermal motion of the atoms as quenched disorder, where all observables are computed as averages over $M$ unitary runs. Atom positions are perturbed randomly before each of the $M$ unitary simulations, but are considered fixed throughout a single run. The expectation values are then computed as averages over the expectation values of the $M$ random realizations. The positional fluctuations are modeled as
\begin{equation}
    \langle \delta r_{x/y, i}^2 \rangle^{1/2} = 0.2 \mu m \ ,
\end{equation}
where the $\delta r_{x/y, i}$ are the components of the error on the position of atom $i$ for a given unitary run. In the same manner, we model the laser imperfections as shot-by-shot fluctuations with a relative root-mean-square (RMS) error of the Rabi frequency
\begin{equation}
    \frac{\langle \epsilon_\Omega^2 \rangle^{1/2}}{\langle \Omega \rangle} = 0.008 \ ,
\end{equation}
as well as fluctuations in detuning $\Delta$ with RMS error
\begin{equation}
    \langle \epsilon_\Delta^2 \rangle^{1/2} = 0.18 \frac{\text{rad}}{\mu s} \ .
\end{equation}
All expectation values of observables are then averaged over $M=100$ realizations of the random perturbations on the atom position, Rabi frequency and detuning. For the RO error, we consider the asymmetric ground and Rydberg state confusion with $P(r|g) = 0.01$ and $P(g|r) = 0.08$, which is modeled as a quantum channel after the unitary evolutions. The error model is explained in more detail in Appendix~\ref{sec:AppendixErrorModel}.

We provide only a minimal error model that qualitatively explains the experimental data. The main contributions to this error model are RO errors~\cite{Wurtz_2023} and atomic motion, as we discuss in Appendix~\ref{sec:AppendixErrorModel}. Atomic motion has previously been identified as the leading noise source~\cite{Dag_2025}. Note that the error model does not require any fitting of free parameters. However, it has been shown that fitting the error model's parameters improves agreement between experimental data and the error model predictions~\cite{Dag_2025}, as well as provides a more accurate picture of the device properties. Although our model explicitly captures the main dephasing channels, we leave several other decoherence channels unconsidered, such as the laser inhomogeneity across the chamber, atom loss during the simulation or $T_1$-decay. More careful error modeling, together with more advanced error mitigation techniques, is left for future research.

However, we implement rudimentary error mitigation by post-selecting on the correct subspace. Since in the ideal Heisenberg XXZ chain the excitation number is conserved, we can discard all shots that do not have the same number of domain walls as the initial state. This condition is easily checked. The post-selection significantly improves the accuracy of the measured observables, but it results in $\sim 50\%$ of shots wasted, and  even up to $\sim 70\%$ wasted for the simulation of $\delta=4.0$. This is consistent with the observation that the error of the domain wall encoding is the most pronounced for $\delta=4.0$, as we discuss in the following section. In Appendix~\ref{sec:AppendixB}, we discuss the ratio of discarded shots and the impact of post-selection on the observables in more detail. Our post-selection also discards shots where Aquila reports an atom loading failure, i.e. an atom position that was supposed to capture an atom was detected to be empty before the pulse sequence was applied. Thus, post-selection also mitigates atom loading errors.

\subsection{Discussion of the parameter choices}
The correspondence between $H_{Ryd}$ in Eq.~\eqref{eq:HRyd} and $H_{XXZ}$ in Eq.~\eqref{eq:HSpinChain} is only exact up to inconsequential added constants in the limit $J \rightarrow \infty$~\cite{Werner_2024}. For finite $J$, there is an error to the Hamiltonian with a worst-case bound of $\mathcal{O}(J^{-1})$, resulting in a bound on the infidelity of the simulation of $\mathcal{O}(t^2J^{-2})$, where $t$ is the simulation time. Here we set $J = 5 / 3 \approx 1.67$. Hence, the larger $J$, the longer the XXZ chain can be simulated accurately. In the numerical investigation below, we see that for the time scales we probe here, our choice of $J$ is sufficiently large.

Although increasing $J$ would improve the quality of the simulation in principle, here we have to balance several trade-offs. Since increasing $J$ requires moving the atoms closer together, it becomes more difficult to find a positioning of the atoms that respects Aquila's hardware constraints, where atoms must be more than $4\mu m$ apart, both radially, as well as along the y-dimension. A similar consideration must be made when choosing the system size $N=13$, since the atoms must fit within a $75\text{$\mu$m}\times76\text{$\mu$m}$ rectangle.

Here, we compute the distances required for the desired nearest-neighbor and next-nearest-neighbor interactions and arrange the atoms in the pattern that respects the necessary angles, as shown in Figure~\ref{fig:register}(a). The shape as a whole is then rotated in the xy-plane until the geometric constraints of Aquila are respected. If $J$ is too large and the atoms are too close, this becomes impossible to solve.

Furthermore, we set the Rabi frequency during the simulation to $\Omega = 1.58 \ \text{rad} \ \mu s^{-1}$ as the global energy scale. In order to simulate a longer time-scale before the system decoheres, it would be prudent to increase $\Omega$. However, to maintain the same $J$, this would also lead to a closer arrangement of the atoms.

At the same time, when the atoms are placed closer together to increase $J$, the sensitivity to position fluctuations, as well as the unwanted interactions beyond the next-nearest neighbor also become stronger. This means that their thermal fluctuations may also affect the simulation more strongly. Similarly, choosing a larger anisotropy $\delta$ will result in a tighter configuration of atoms, which might negatively impact simulation fidelity. This would limit the range of $\delta$ accessible at an acceptable drop in simulation quality. Furthermore, $4J \gg \delta$ is required to maintain the energetic separation of subspaces of distinct domain wall number, which allows for the interpretation of the domain walls as particles evolving in time. This condition is only moderately satisfied for $\delta = 4.0$, which explains why the fidelity degrades the most for this value of $\delta$, as we discuss in Section~\ref{sec:Numerics}.

However, note that in our approach there is no hard ceiling to $\delta$ as is found in approaches based on Floquet engineering~\cite{Scholl_2022} and we can probe $\delta=4.0$ with reasonable accuracy, with deviations explained by our noise model. This is twice as large as the anisotropies reported in Ref.~\cite{Scholl_2022}, where $\delta \leq 2$ due to constraints of low-order Floquet engineering of the native XY interactions.

Hence, we have to choose the parameters $N$, $\delta$, $\Omega$ and $J$ to balance the domain wall error, thermal fluctuations, geometric hardware constraints and decoherence. Here, we hand-picked a set of parameters that works sufficiently well to probe the phases of the XXZ model. However, it is likely that an extensive optimization of the Rabi frequency and atom distances can improve the quality and size of the quantum simulation presented here. We leave this investigation for future research.

\subsection{Domain wall encoding of observables}
It is worth highlighting here that in the following, we will consider the domain wall encoded operators of observables~\cite{Werner_2024}. Since domain walls live on the bonds between atoms, the domain wall Pauli-Z operator on site $i$ is given by
\begin{equation}
    \begin{aligned}
        \sigma^z_{DW,1} &= -\sigma_1^z \\
        \sigma^z_{DW,i} &= \sigma_{i-1}^z \sigma_i^z \  \ \text{if} \ 1 < i < N \ , \\
        \sigma^z_{DW,N} &= -\sigma_{N-1}^z \ ,
    \end{aligned}
\end{equation}
where the sign of the edge cases is determined by the orientation of the virtual spins. The site occupation operators are then
\begin{equation}
    \hat{n}_{DW,i} = \left( 1 + \sigma^z_{DW,i} \right) / 2 \ .
\end{equation}
In the following, we consider domain wall operators exclusively, except when comparing observables to the exact model. From context, it should be clear whether we consider an observable in the exact model, the domain-wall encoded version in the Rydberg simulator, or the numerical simulation of the Rydberg system. Therefore, we will omit the subscript ``DW'' for clarity.
\section{Numerical results}
\label{sec:Numerics}
\begin{figure}
    \centering
    \includegraphics[scale=.3875]{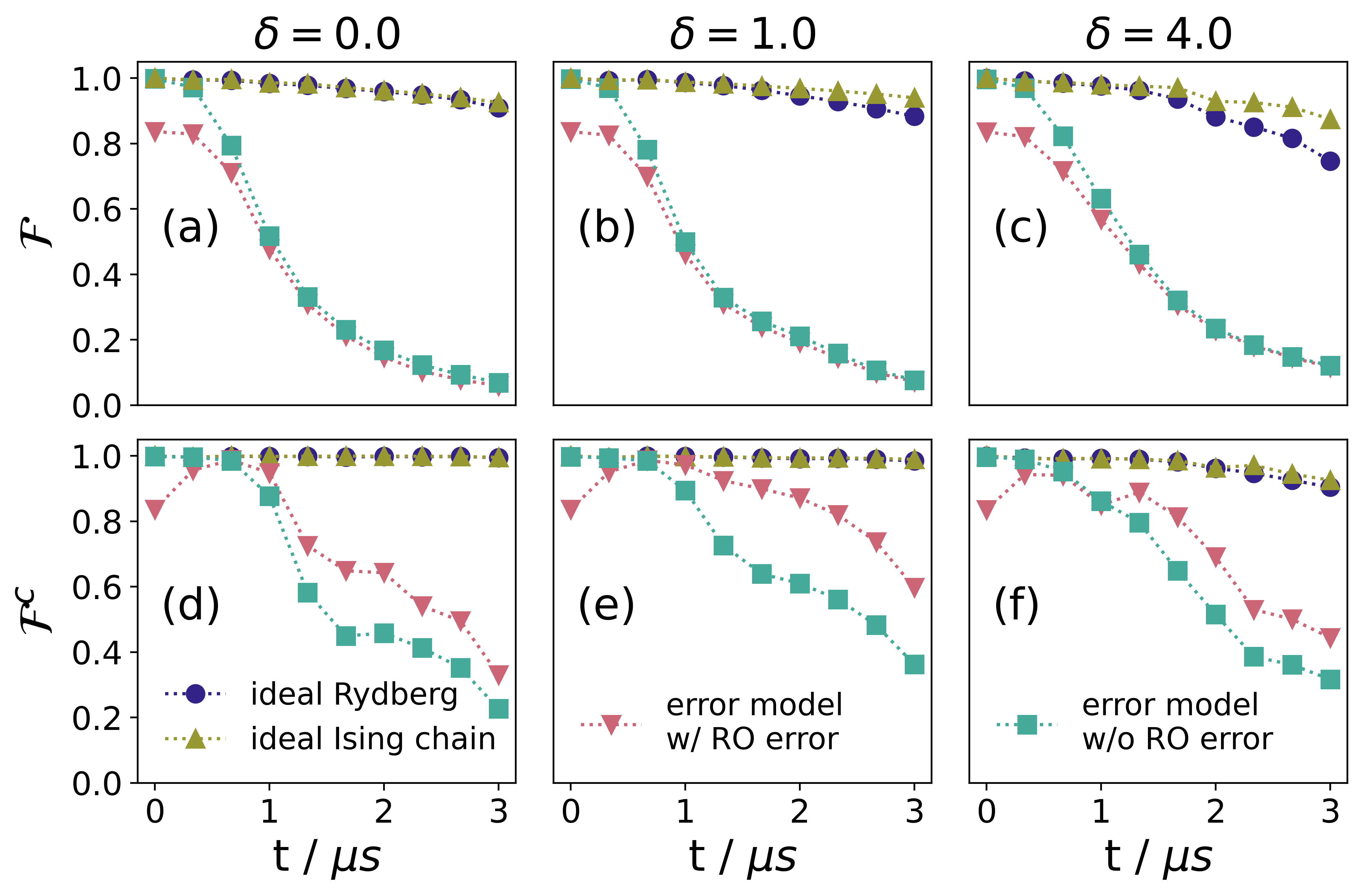}
    \caption{Fidelities of an ideal Rydberg simulation (blue dots), for the full error model (red triangles) and for the same error model, but without RO error (green squares) compared to the exact Heisenberg XXZ evolution in the XY phase (a, d), at the critical point (b, e) and in the Ising phase (c, f). Additionally, we show the fidelities for an ideal Ising chain, i.e. the Hamiltonian of the Rydberg system where the unwanted interactions are set to zero (yellow upward-triangles). (a, b, c): the quantum state fidelity $\mathcal{F}$ in Eq.~\eqref{eq:Fidelity}; (d, e, f): the classical fidelity $\mathcal{F}^c$ in Eq.~\eqref{eq:ClassicalFidelity}.}
    \label{fig:Fidelities}
\end{figure}
In order to validate our choices of parameters, as well as to break down the contribution to the error of the simulation, we numerically compute the state fidelity with the exact Heisenberg XXZ evolution for the time-evolved full error model, the time-evolved error model only with atomic disorder and laser imperfections, but without RO error, and a simulation of the ideal Rydberg simulator, which does not suffer from atom motion, laser imperfections, or RO error, but realizes $H_{Ryd}$ exactly. Moreover, we compute the fidelities of the time evolution under an idealized Ising chain, which is the same idealized system and Hamiltonian $H_{Ryd}$ as for the ideal Rydberg simulation, but where we additionally remove the unwanted interactions beyond next-nearest neighbor. Note that the ideal Rydberg simulation still includes the errors due to finite domain wall coupling $J$ and unwanted interactions of the Rydberg atoms beyond next-nearest-neighbor, while the ideal Ising chain only suffers from the $\mathcal{O}(J^{-1})$ error of the domain wall encoding. For all models, we apply the post-selection as discussed in more detail in Appendix~\ref{sec:AppendixB}. More mathematical details of the implementation of the numerical simulation can be found in Appendix~\ref{sec:AppendixPostselection}.

We compute the respective fidelities with the time evolution of the true XXZ chain $\ket{\psi(t)} = \exp(-itH_{XXZ}) \ket{\psi(0)}$, i.e.
\begin{equation}
    \mathcal{F}(t) = \bra{\psi(t)} \rho(t) \ket{\psi(t)} \ .
    \label{eq:Fidelity}
\end{equation}
We additionally compute the classical fidelity
\begin{equation}
    \mathcal{F}^c(t) = \left( \sum_z |\langle z | \psi(t)\rangle|\sqrt{\bra{z} \rho(t) \ket{z}} \right)^2 \ .
    \label{eq:ClassicalFidelity}
\end{equation}
The classical fidelity measures the fidelity of the probability distribution in the computational basis, i.e. it makes it possible to gauge the quality of the simulation in reproducing diagonal observables. Since the device limits us to computational basis measurements, all accessible observables are diagonal.

The results are shown in Figure~\ref{fig:Fidelities}. We see that the fidelities in Figure~\ref{fig:Fidelities}(a, b, c) under the error model with atom motion decay rapidly over time, both with and without considering the RO error. In contrast, the fidelity of the ideal Rydberg system without atom motion or RO error remains high (above 0.75) throughout the entire evolution for all $\delta$ and even higher for $\delta = 0.0$ and $\delta=1.0$ in Figure~\ref{fig:Fidelities}(a, b). Interactions beyond next-nearest-neighbor begin to introduce slightly more errors as $\delta$ grows, as seen by the increased distance between the fidelities of the ideal Rydberg and the ideal Ising simulation in Figure~\ref{fig:Fidelities}(c) for $\delta=4.0$. However, dropping the unwanted interactions to simulate the ideal Ising chain only improves the fidelity slightly for any $\delta$, indicating that the unwanted interactions are indeed close to negligible and that the observed errors in the ideal Rydberg atom simulation are dominated by the controllable error of the domain wall encoding. This error affects the simulation of $\delta=4.0$ the most, as for a fixed value of $J$, the energetic separation between subspaces of distinct domain wall number is the smallest.

Although the RO error degrades the quantum simulation slightly, among the considered error modalities, the main contribution is due to the thermal fluctuations of the atom positions, which translates into a per-shot noise of the Hamiltonian parameters. This induces disorder, which inhibits the ballistic motion of the excitations, as has been previously identified in Ref.~\cite{Dag_2025}. The contributions of the laser imperfections to the total error are negligible, as discussed in Appendix~\ref{sec:AppendixErrorModel}.

These numerical results indicate that, in principle, the domain wall encoding applied to an idealized Rydberg atom simulator can reproduce the dynamics of the XXZ chain with reasonable accuracy, as indicated by the fidelity consistently above 0.75 for the entire simulation time interval in Figure~\ref{fig:Fidelities}(a, b, c). This is even the case here, where $J=5/3$ is quite small due to balancing the different constraints of the device. Once the thermal motion of the atoms is considered, the fidelity drops significantly. This additional error could potentially be mitigated by improved engineering, such as better cooling of the atoms.

At the same time, the classical fidelity does not decay as fast, even when thermal motion is introduced, so diagonal observables can potentially be better reproduced qualitatively, as shown in Figure~\ref{fig:Fidelities}(d, e, f). This is expected, since the classical fidelity $\mathcal{F}^c$ is lower bounded by the fidelity $\mathcal{F}$. Interestingly, while the RO error seems to have a detrimental effect on the fidelity for short times and a negligible effect for later times when other error sources become dominant, it seems to improve the classical fidelity at later times. This improvement is quite substantial when comparing Figure~\ref{fig:Fidelities}(a, b, c) with (d, e, f).

We do not have a conclusive explanation for this observation, but we speculate that it could be explained by the fact that RO errors of qubits adjacent to a domain wall do not get discarded by the post-selection and effectively move the domain wall. This partially counteracts the inhibited motion of the excitation by the thermal fluctuations of the atoms. However, one would expect this effect to also improve the fidelity $\mathcal{F}$, but it could be that the advantage here is overruled by the general decay in $\mathcal{F}$.

The RO error on top of the thermal atom motion indicates the maximum improvement that can be expected from introducing RO error mitigation techniques alone, which is a moderate improvement in this case. The main improvement of the simulation can be expected from improved cooling of the atoms.

All numerical simulations were implemented using QuTiP~\cite{Qutip1, Qutip2} and Amazon Braket SDK~\cite{braket2019}. For the error model, we average all results over $M=100$ realizations of the positional and laser parameter disorder.

\section{Experimental results}
\begin{figure}
    \centering
    \includegraphics[scale=.45]{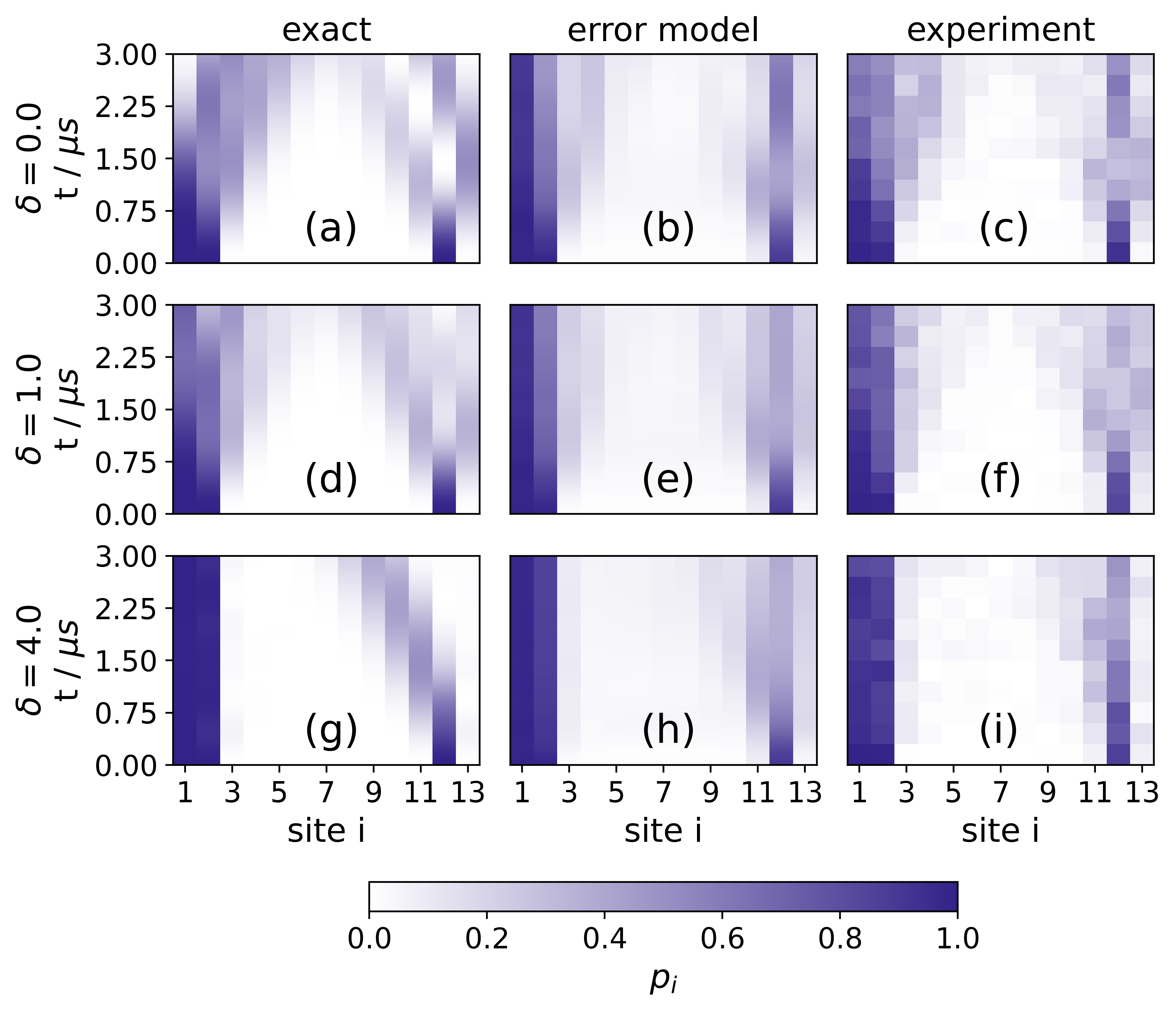}
    \caption{Time evolution of the site populations $p_{i}$ Eq.~\eqref{eq:OnSitePopulations} as a function of time $t$ for the anisotropic XXZ chain with a total of three excitations, two initialized at sites 1 and 2 and one at site 12. We compare the exact solution (a, d, g) with the numerical simulation of the error model (b, e, h) and the experimental data from Aquila (c, f, i) for the system in the XY phase $\delta=0.0$ (a, b, c), at the critical point $\delta = 1.0$ (d, e, f) and in the Ising phase $\delta=4.0$ (g, h, i).}
    \label{fig:XXZ_population}
\end{figure}

\begin{figure*}
    \centering
    \includegraphics[scale=.3375]{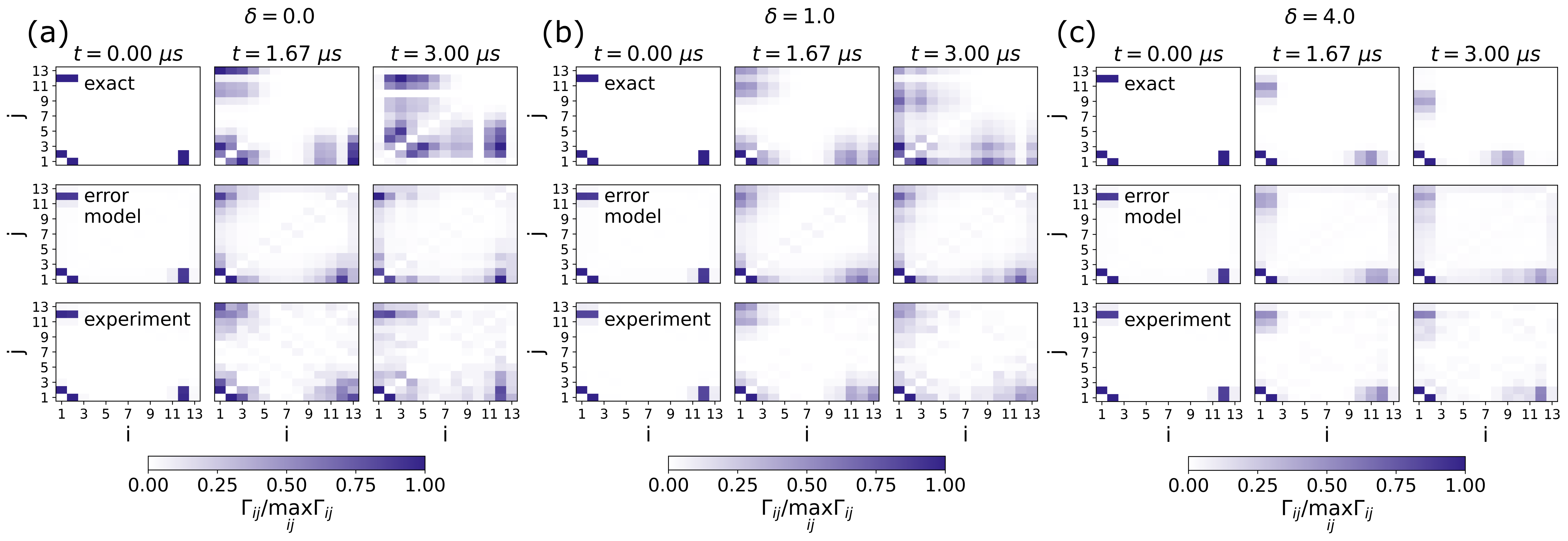}
    \caption{Time evolution of the normalized correlation matrix $\Gamma_{ij} / \max_{ij} \Gamma_{ij}$ for the exact model (top row), the error model (middle row) and the experimental data (bottom row) for the XY phase with $\delta=0.0$~(a), the critical point with $\delta=1.0$~(b) and in the Ising phase with $\delta=4.0$~(c). In each subfigure, we show the normalized correlation matrix at the start of the simulation at $t=0.0 \mu s$ (left column), in the approximate middle of the simulation $t=1.67 \mu s$ (middle column) and at the end of the simulation at $t=3.0 \mu s$ (right column).}
    \label{fig:correlations}
\end{figure*}
We perform the quantum simulation on Aquila for 300 shots per time point. First, we investigate the on-site populations
\begin{equation}
    p_i = \langle \hat{n}_i \rangle \ ,
    \label{eq:OnSitePopulations}
\end{equation}
shown in Figure~\ref{fig:XXZ_population} for the exact Heisenberg model (a, d, g), the error model (b, e, h) and the experimental data obtained from Aquila (c, f, i) for $\delta=0.0$ (a, b, c), $\delta=1.0$ (d, e, f) and $\delta=4.0$ (g, h, i). Visually, the error model and the experimental data seem to agree, but both deviate from the exact simulation. This can be mainly attributed to the thermal atom motion. Since the RO error is time-independent and does not affect the dynamics, it does not explain the reduction in motion of the excitations, but rather explains the blurring of the heatmaps.

However, qualitatively, the dynamics are reasonably well reproduced by the experiment. The key difference between the XY and the Ising phase is the formation of stable bound states for excitations. We initialize the system with a single excitation on one end, and two adjacent excitations on the other. In the exact model, the individual excitation moves freely in the bulk for all values of $\delta$, with the exception that it scatters off the virtual fixed excitation on site $N+1$ for $\delta \neq 0$. Note that the virtual fixed excitation at site $N+1$ is a consequence of the local boundary field. In the XY phase, the adjacent excitations bind loosely, or not at all at $\delta=0.0$, which is why they move apart. In the Ising phase, however, the adjacent excitations form a bound state that corresponds to a single particle with a much larger effective mass, since now the motion of the bound particle requires a second-order effect, which is suppressed by increasing $\delta$. Additionally, since the excitations are located at the end of the chain, they also bind to the virtual excitation at site $i=0$, i.e. the logical boundary field. The contrast between the freezing of the bound particles and the maintained motion of the individual excitation can be observed in the experimental data as well, including the scattering off the virtual excitation at site $N+1$.

Additionally, we measure the two-site correlator~\cite{Kim_2024}
\begin{equation}
	\Gamma_{ij} = \langle \hat{n}_i \hat{n}_j \rangle \ .
\end{equation}
The results are shown in Figure~\ref{fig:correlations} for three distinct points in time. For clarity, we normalize the correlation matrix by the observed maximum $\max_{ij} \Gamma_{ij}$.

Figure~\ref{fig:correlations} clearly shows how in the exact model the two-body correlations spread through the chain in the XY phase shown in (a), while remaining localized in the Ising phase shown in (c). At the critical point shown in (b), there is only a moderate spread of correlations. The spread of correlations in the experimental data seems to be inhibited compared to the exact model, though we see a clear difference between the XY and Ising phase. The reduction in spread seems to be explained by the error model.

\begin{figure}
    \centering
    \includegraphics[scale=.475]{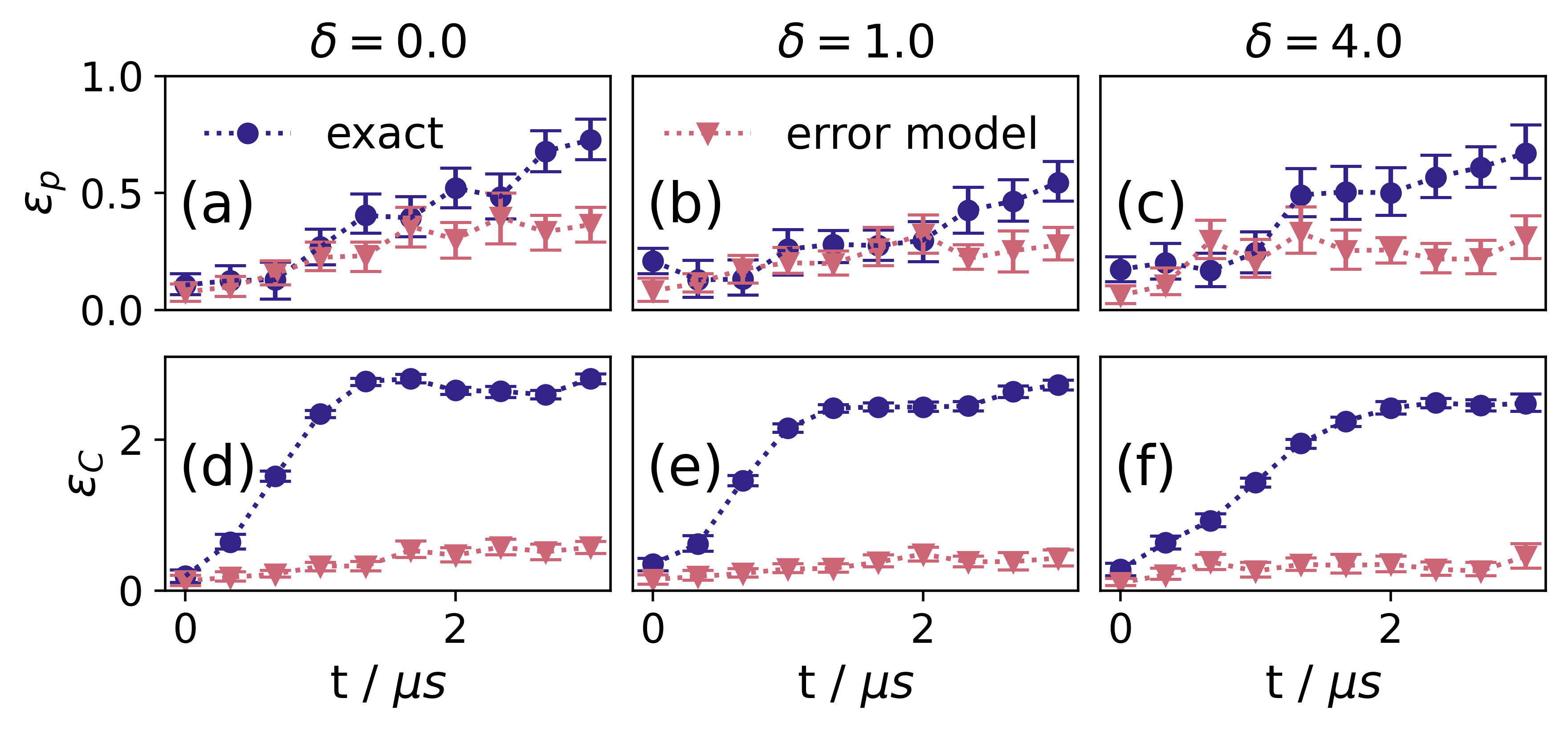}
    \caption{Error metrics $\varepsilon_p$ Eq.~\eqref{eq:EpsP} (a, b, c), and $\varepsilon_C$ Eq.~\eqref{eq:EpsC} (d, e, f) for the experimental data being compared to once the exact time evolution (blue dots) and once the error model (red triangles) for $\delta = 0.0$ (a, d), $\delta = 1.0$ (b, e) and $\delta = 4.0$ (c, f). The markers are the mean values from bootstrapping the experimental data 100 times and the error bars indicate the 95\%-confidence interval.}
    \label{fig:error_metrics}
\end{figure}

Both Figure~\ref{fig:XXZ_population} and Figure~\ref{fig:correlations} show that the experimental data appear to match the predictions of the error model better than the exact model's predictions. This observation from visual inspection can be made concrete by considering the covariances
\begin{equation}
    C_{ij} = \langle \hat{n}_i \hat{n}_j \rangle - \langle \hat{n}_i \rangle \langle \hat{n}_j \rangle
\end{equation}
and computing the errors
\begin{equation}
    \varepsilon_p = \left( \sum_i ( p_i  -  \tilde{p}_i )^2 \right)^{\frac{1}{2}} \ ,
    \label{eq:EpsP}
\end{equation}
and
\begin{equation}
    \varepsilon_C = \left( \sum_{ij} (C_{ij} - \tilde{C}_{ij})^2 \right)^{\frac{1}{2}}
    \label{eq:EpsC}
\end{equation}
where the tilde indicates the respective quantity derived from the numerical simulation. We compare the experimental data once to the numerical simulation of the exact XXZ chain and once to the error model. We use these error metrics based on one- and two-local observables since they are known to be more robust in the low-shot-number regime~\cite{King_2025}. We show the evolution of the errors $\varepsilon_p$ and $\varepsilon_C$ as functions of time in Figure~\ref{fig:error_metrics}, both for the exact XXZ model and the error model as the numerical comparison. From this, it is clear that the error model explains the experimental data much better, while the difference is the most significant for the covariance error.

As a final benchmark, we compute the observable
\begin{equation}
	N_{\text{flip}} = \frac{1}{2} \sum_{i=1}^N \left( 1 - \langle \sigma^z_i \sigma^z_{i+1} \rangle \right) \ ,
    \label{eq:Nflip}
\end{equation}
which measures the number of unaligned neighboring spins, or logical domain walls, as a function of evolution time~\cite{Scholl_2022}. Note that here we mean the logical domain walls in the simulated XXZ spin chain, not the physical domain walls in the Ising model, which correspond to excitations in the spin chain.

\begin{figure}
    \centering
    \includegraphics[scale=.3875]{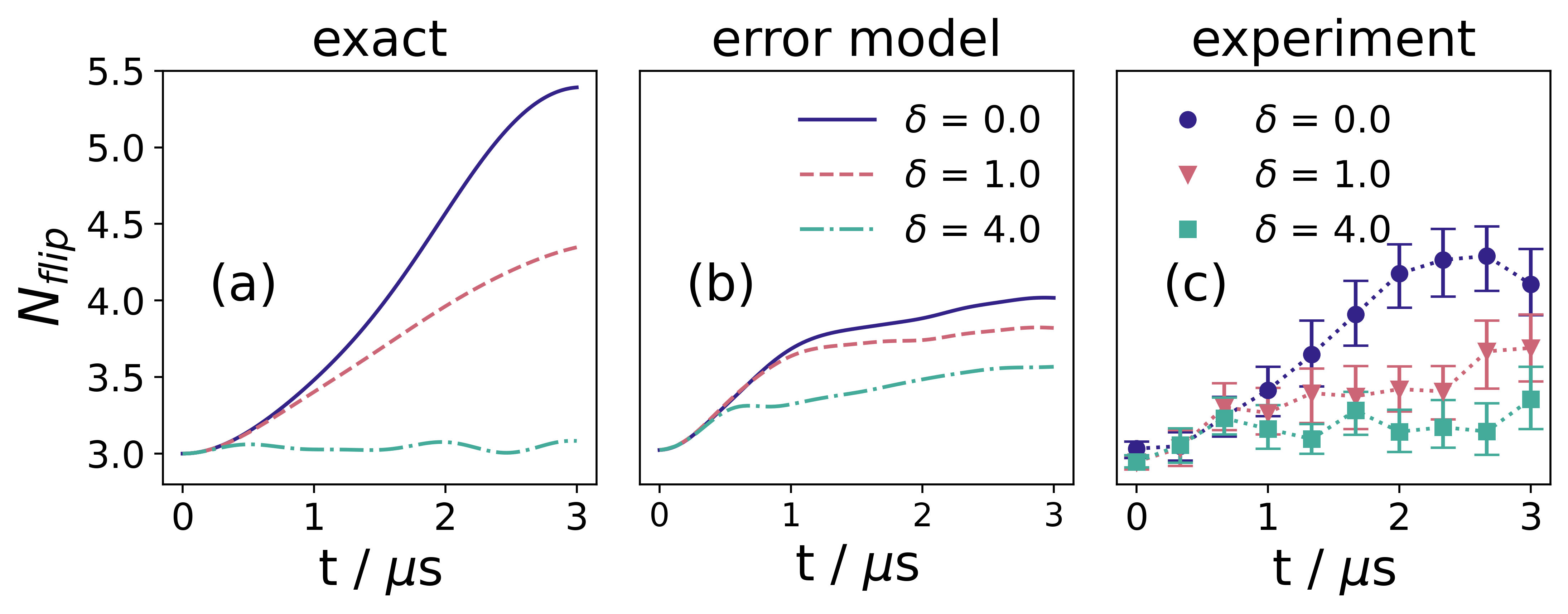}
    \caption{$N_{\text{flip}}$ from Eq.~\eqref{eq:Nflip} computed for the exact model (a), the error model (b) and using the experimental data (c) for all values of $\delta$. In (c), the markers are the mean values from bootstrapping the experimental data 100 times and the error bars indicate the 95\%-confidence interval.}
    \label{fig:NflipResult}
\end{figure}

In the Ising phase, $N_{\text{flip}}$ is an approximately conserved quantity, since the large anisotropy $\delta$ energetically inhibits the formation or dissolution of bound excitations. Hence, the number of unaligned adjacent spins does not change significantly over time for large $\delta$. This effect is visible in the exact XXZ simulation, the error model and the experimental data, as seen in Figure~\ref{fig:NflipResult}. For all three simulations, we can see that $N_{\text{flip}}$ remains close to constant in the Ising phase ($\delta=4.0$) and grows in the XY phase ($\delta=0.0$). At the critical point, $N_{\text{flip}}$ also grows, but only to an intermediate value.

This behavior is consistent with the exact model shown in Figure~\ref{fig:NflipResult}(a), the error model shown in (b) and the experimental data shown in (c), while the growth in the experimental quantum simulation is inhibited despite the qualitative order being maintained. From comparison with the error model, we can see that the error model explains a much slower growth of $N_{\text{flip}}$. This is consistent with the observed reduction in motion as a result of fluctuations in the atom positions. The experimental results for the XY and Ising phases match the error model predictions, while $N_{\text{flip}}$ at the critical point grows slightly slower than predicted.

In total, the experimental data maintain the predicted ordering of the growth of $N_{\text{flip}}$ depending on the anisotropy $\delta$. Thus, the experiment reproduces the exact model qualitatively, with the quantitative deviations explained by the error model.
\section{Conclusion and outlook}
We have applied the method of domain wall encoding on an Ising chain with nearest- and next-nearest-neighbor ZZ-interactions to simulate the one-dimensional Heisenberg XXZ model on an array of Rydberg atoms. This allowed us to simulate the XXZ chain both in the XY and the Ising phase, as well as at the critical point, and qualitatively reproduce the distinct behavior of physically relevant one- and two-body observables in each phase.

Our numerical investigations have shown that the domain wall mapping reproduces the Heisenberg XXZ model with high fidelity (above 0.75 for all $\delta$), even if the errors due to finite domain wall coupling and undesired long-range interactions are considered. The main source of error occurs due to imperfections of the real hardware. Here we have considered thermal fluctuations of the atom positions as the main source of error, which explains qualitatively the experimentally observed data. Therefore, we expect that improved cooling would greatly enhance the quantum simulation discussed here.

Although our error model qualitatively explains the experimental observations, as measured by the relevant single-body and two-body observables, we do not claim that it gives a full accounting of all error sources. While we model RO error, laser imperfections, and fluctuations of the atom positions and account for atom loading failures identified before executing the pulse sequence by discarding the respective shots, we neglect other decoherence channels. Specifically, we do not consider atom loss during the simulation and the inhomogeneity of the Rabi frequency across the chamber, nor decoherence due to decay of the Rydberg state. A more detailed consideration of all error sources may explain the remaining discrepancy between our error model and the observed data.

Our method allowed us to simulate Heisenberg XXZ chains outside the single-excitation sector with anisotropies beyond the hard algebraic ceiling of low-order Floquet engineering of the native XY interactions~\cite{Scholl_2022}. Furthermore, to our knowledge, we report the first implementation of our method to simulate longitudinal interactions of excitations.

In principle, domain wall encoding allows us to realize also disordered systems~\cite{Werner_2024}, which can feature a much richer phase diagram than the homogeneous Heisenberg XXZ. To realize these models on Rydberg atom arrays, it is required to overcome the limitations of the local addressing. Here, we used the local detuning to prepare the system in the desired initial state. However, due to the limitations in the time-dependent control of the local detuning, the feature cannot be used to, for example, compensate for local fields that arise when transforming from occupation number $\hat{n}$ to $\sigma^z$, unless there is a correspondence between the required local fields and the initial state, which would be a strong assumption. Additional control of the time dependence of the local detuning would greatly enhance the capabilities of the device to simulate various interesting models.

Similarly, the homogeneity of the driving laser -- and thus the Rabi frequency -- limits us to spin chains with homogeneous hopping energies. Using a second driving term, one could also simulate a topological model like the Su-Schrieffer-Heeger model~\cite{Werner_2024}. Recent results on dual-species Rydberg atom simulators~\cite{Anand_2024} are a promising direction for future applications of domain wall encoding for spin chain simulations. Furthermore, one could investigate exploiting the inhomogeneity of the Rabi frequency to generate inhomogeneous hopping energies.

\section*{Acknowledgements}
We thank Ana Palacios for helpful discussions. The Aquila quantum processor was accessed via Amazon Braket. The authors gratefully acknowledge RES resources provided by Barcelona Supercomputing Center in MareNostrum 5 to INNO-2026-1-0004.

\appendix
\section{Error model}
\label{sec:AppendixErrorModel}
We consider three main error sources, namely the thermal motion of the atoms, the RO error and laser imperfections. This models the main contributions to the dephasing rates of the system.

Thermal fluctuations of atoms have been shown to inhibit excitations traveling in an Ising chain, where ballistic transport is suppressed~\cite{Wang_2025, Dag_2025}. As we show in the main text, our error model captures the relevant physics. Since this is sufficient to explain our observations, the calibration data reported in Ref.~\cite{Wurtz_2023} might be overly cautious or outdated due to system upgrades and fitting the model parameters might result in more accurate values.

We model the atom motion as quenched disorder, i.e. we introduce errors to the atom positions in our numerical simulations. We assume the atoms to be located in the xy-plane and add random perturbations to the atom positions such that
\begin{equation}
	\vec{r}_i \rightarrow \tilde{\vec{r}}_i = \vec{r}_i + \delta \vec{r}_i \  ,
\end{equation}
where $\vec{r}_i \in \mathbb{R}^2$ is the desired position of atom $i$, while $\tilde{\vec{r}}_i$ is the actual position, which differs from the target position by an error where the components $\delta \vec{r}_i = (\delta r_{x,i}, \delta r_{y,i})^T$ are random numbers
\begin{equation}
	\delta r_{x/y, i} \sim \mathcal{N} \left( 0, (0.2 \mu m)^2 \right) \ .
\end{equation}
This models a deviation of $0.2 \mu m$ in both directions around the desired location $\vec{r}_i$. The magnitude of the fluctuations is taken from Ref.~\cite{Wurtz_2023}. During the unitary evolution, the positional disorder is kept fixed.

Similarly, we model the laser imperfections. Before simulating each unitary evolution, we replace the Rabi frequency $\Omega$ with
\begin{equation}
    \Omega \rightarrow \tilde{\Omega} = (1+\epsilon_\Omega) \Omega \ ,
\end{equation}
where
\begin{equation}
    \epsilon_\Omega \sim \mathcal{N}(0, 0.008^2) \ .
\end{equation}
For the detuning $\Delta$ we set for each run
\begin{equation}
    \Delta \rightarrow \tilde{\Delta} = \Delta + \epsilon_\Delta \ ,
\end{equation}
with
\begin{equation}
    \epsilon_\Delta \sim \mathcal{N}(0, 0.18^2) \ .
\end{equation}
All observables are averaged over 100 realizations of $\tilde{\Omega}$, $\tilde{\Delta}$ and $\tilde{\vec{r}}_i$.

The last error source we consider here is the misclassification error, where an atom in the Rydberg state is erroneously classified as $\ket{g}$ with probability $P(g|r)$ and vice versa for $P(r|g)$. The misclassification error is not symmetric and is given by $P(g|r)=0.08$ and $P(r|g) = 0.01$~\cite{Wurtz_2023}. In the numerical model, this error is represented by applying single-qubit quantum channels after the unitary evolution. Each single-qubit channel is given by
\begin{equation}
    D_i[\rho] = K_{1,i} \rho K_{1,i}^\dagger + K_{2,i} \rho K_{2,i}^\dagger + K_{3,i} \rho K_{3,i}^\dagger
\end{equation}
where the subscript $i$ indicates that the single-qubit Kraus operators act on qubit $i$. The Kraus operators are
\begin{equation}
    \begin{aligned}
        K_1 &= \begin{pmatrix} 0 & \sqrt{P(g|r)} \\
        0 & 0\end{pmatrix} \ , \\ 
        K_2 &= \begin{pmatrix} 0 & 0 \\
        \sqrt{P(r|g)} & 0\end{pmatrix} \ , \\
        K_3 &= \begin{pmatrix}
            \sqrt{1-P(r|g)} & 0 \\
            0 & \sqrt{1-P(g|r)}
        \end{pmatrix} \ .
    \end{aligned}
\end{equation}
The full RO error is then the composition of all single-qubit error channels
\begin{equation}
    D_{RO}[\rho] = D_1 D_2 ... D_{N-1}[\rho] \ .
\end{equation}
Note that this error model is constructed \textit{a priori} using independently reported calibration data~\cite{Wurtz_2023}, and we do not fit any model parameters to the experimental observations.
\begin{figure}
    \centering
    \includegraphics[scale=.3875]{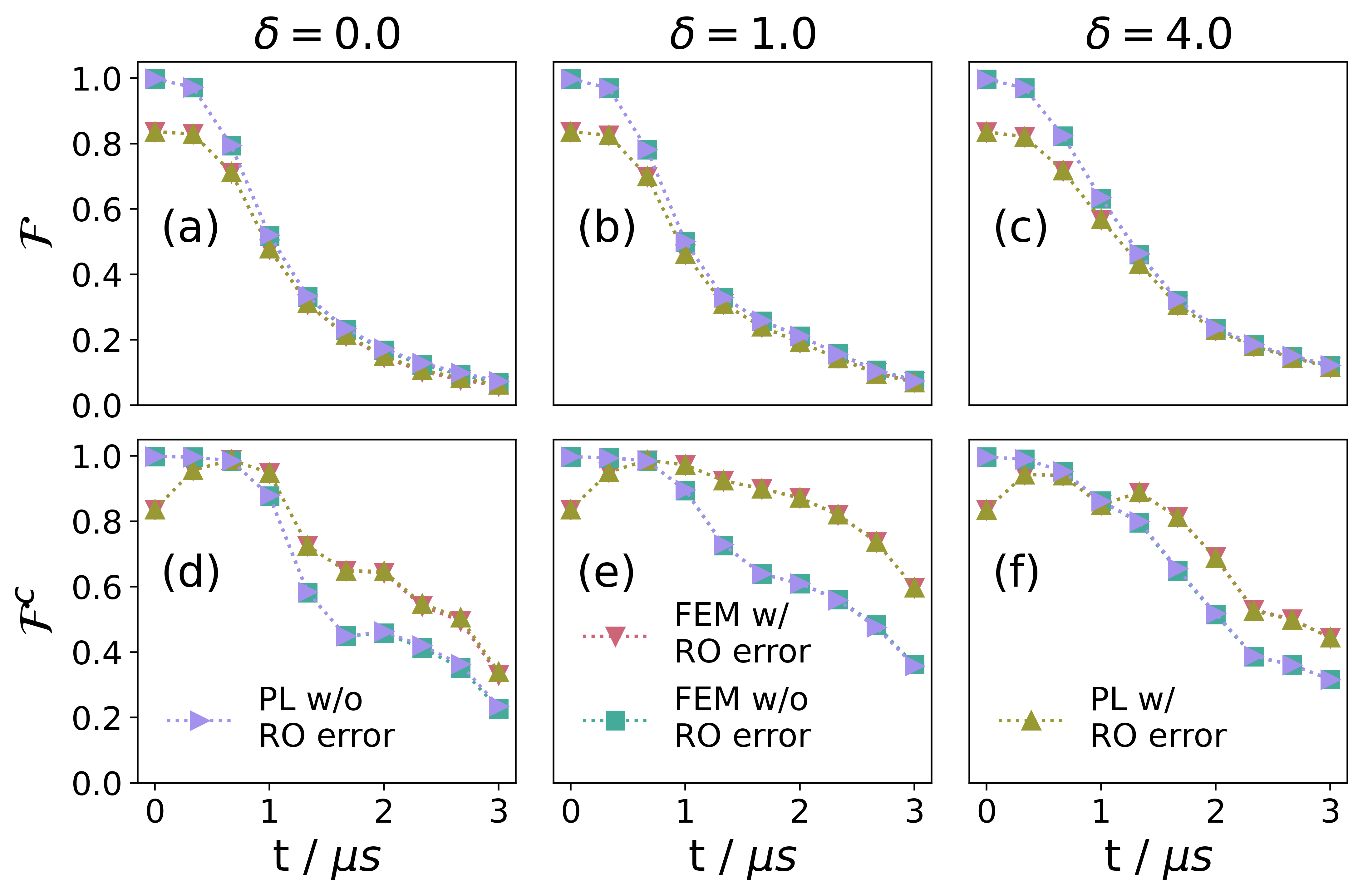}
    \caption{Numerical simulation of the fidelity (a, b, c) and classical fidelity (d, e, f) between the different error models and the exact simulation of the Heisenberg XXZ chain with anisotropy $\delta=0.0$ (a, d), $\delta=1.0$ (b, e) and $\delta=4.0$ (c, f). We consider the full error model (FEM) with and without RO error, as used in the main text, and compare it to the same models but for a perfect laser (PL), where we do not include laser imperfections such as fluctuations in the detuning and the Rabi frequency. We can see that both fidelities are not significantly changed by dropping the laser noise and the RO error contributions are significantly stronger. Note that post-selection is applied in all cases.}
    \label{fig:Fidelities_perfect_laser}
\end{figure}

In Figure~\ref{fig:Fidelities_perfect_laser} we directly compare the state fidelity and the classical fidelity with the exact time evolution of the XXZ chain, once for the full error model including laser imperfections, and once the same error model, but with a perfect laser. We compute both cases once with RO error and once without. We can see that the inclusion of the laser imperfections does not affect the fidelities in a meaningful way for any $\delta$ we test here. Hence, the drop in fidelity between the error model and the ideal Rydberg system discussed in the main text can be mainly attributed to the thermal motion of the atoms and the RO error.

\section{Raw data and post-selection}
\label{sec:AppendixB}
Here, we discuss the ratio of discarded shots due to the post-selection.
\begin{figure}
    \centering
    \includegraphics[scale=.375]{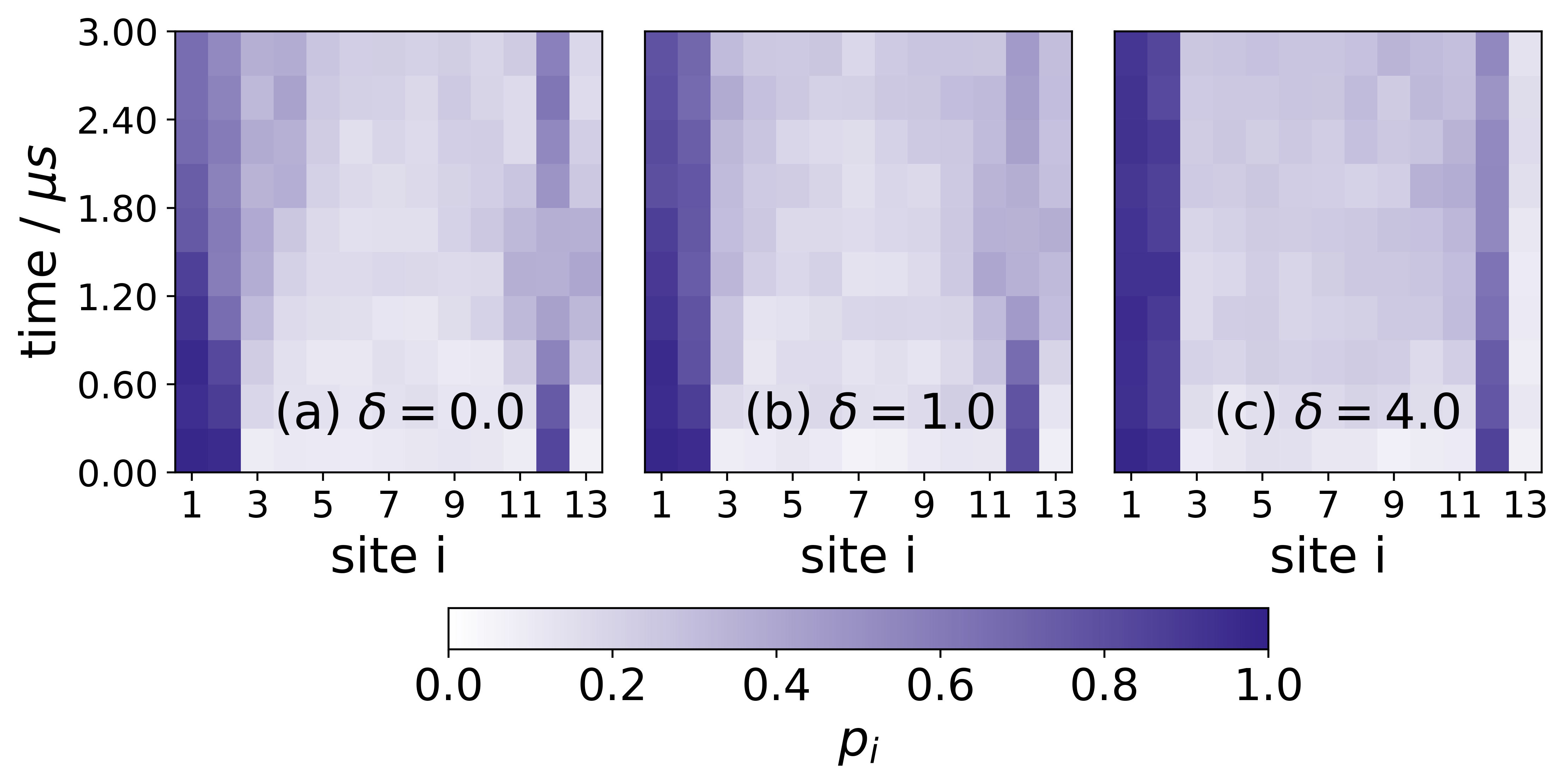}
    \caption{Experimental data of the time evolution of the site populations $p_{i}$ as in Figure~\ref{fig:XXZ_population}, but without applying the post-selection step first for $\delta=0.0$ (a), $\delta=1.0$ (b) and $\delta=4.0$ (c).}
    \label{fig:PostselectStats}
\end{figure}

\begin{figure}
    \centering
    \includegraphics[scale=.46]{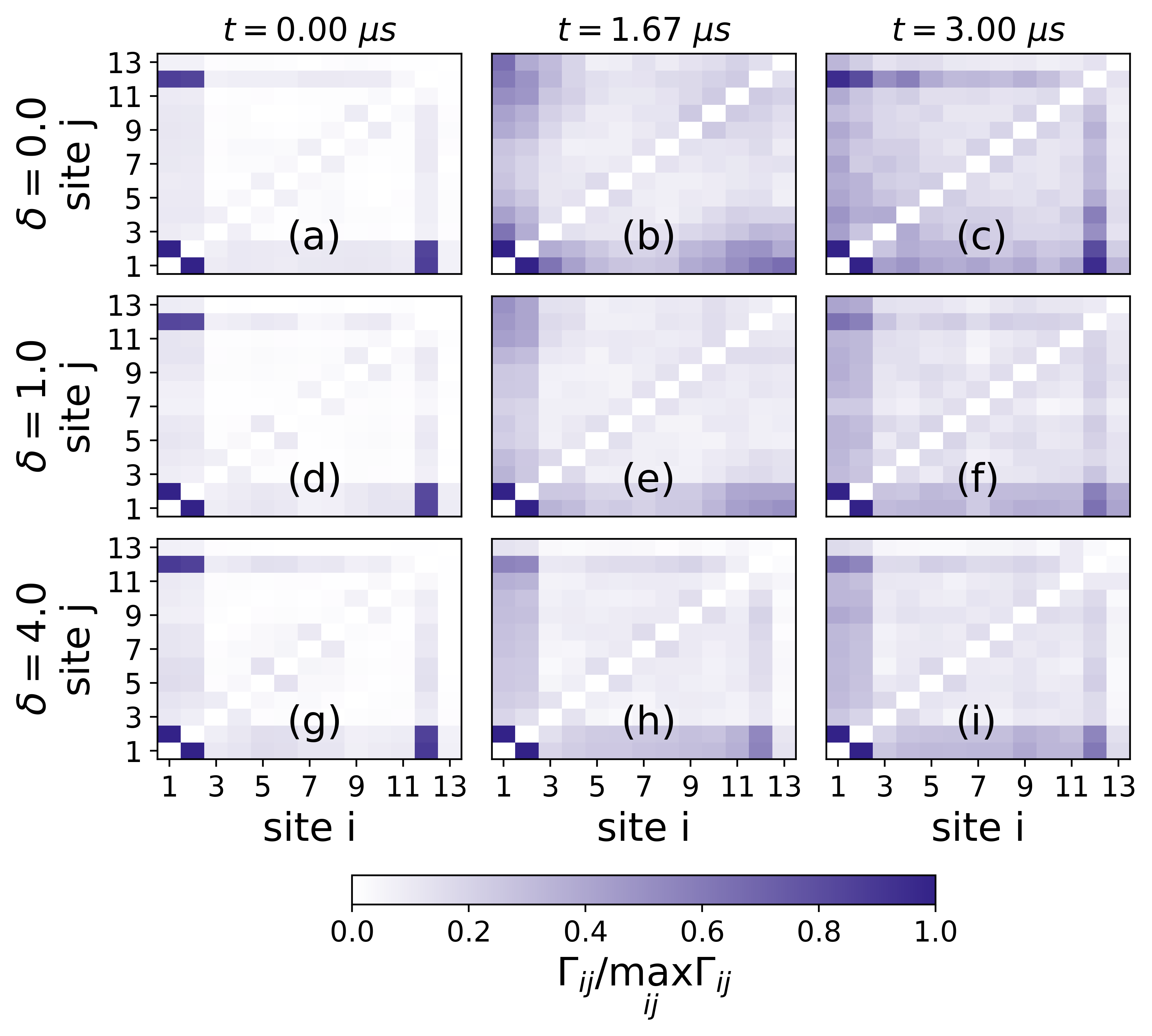}
    \caption{Normalized correlations $\Gamma_{ij} / \max_{ij} \Gamma_{ij}$ at three points in time as in Figure~\ref{fig:correlations} computed from the experimental data without post-selection for $\delta=0.0$ (a, b, c), $\delta= 1.0$ (d, e, f) and $\delta=4.0$ (g, h, i).}
    \label{fig:PostselectCorrelations}
\end{figure}

\begin{figure}
    \centering
    \includegraphics[scale=.45]{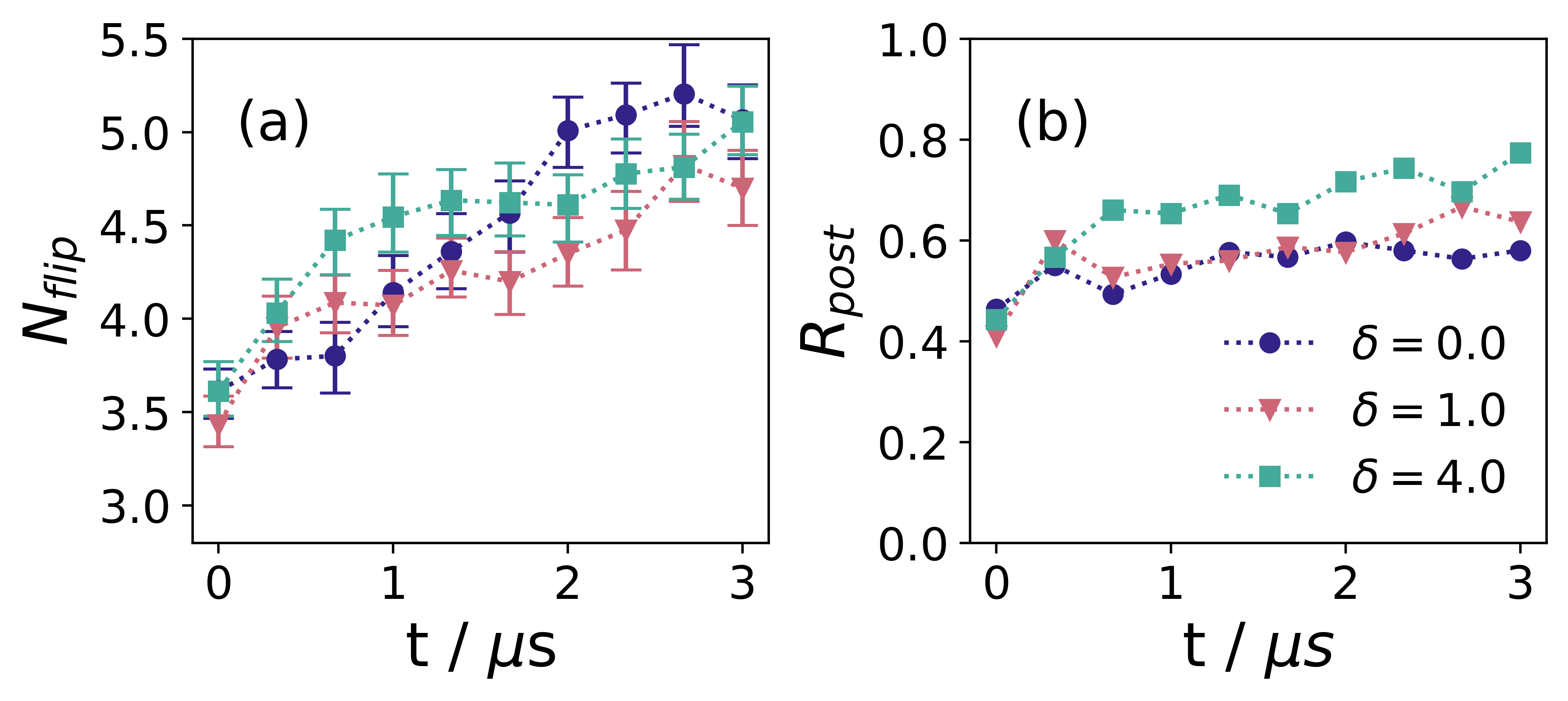}
    \caption{(a): the observable $N_{\text{flip}}$ Eq.~\eqref{eq:Nflip} as in Figure~\ref{fig:NflipResult}, but without post-selection. The markers are the mean values from bootstrapping the experimental data 100 times and the error bars indicate the 95\%-confidence interval. (b): fraction $R_{\text{post}}$ of shots that are discarded by the post-selection on shots with the correct number of excitations and without atom loading errors upon initialization. In both figures, the data is shown for $\delta = 0.0$ (blue dots), $\delta=1.0$ (red triangles) and $\delta=4.0$ (green squares).}
    \label{fig:PostselectNFflip}
\end{figure}

As discussed in the main text, we post-select the data points from Aquila that correspond to the correct number of excitations. This is possible since the model we simulate conserves the number of excitations in the spin chain. Therefore, we can discard shots where we measure a number of domain walls / excitations different from the initial state. Here, we show the effect of this post-selection.\\
In Figure~\ref{fig:PostselectStats}, we show the population as functions of time. We find that the post-selection decreases the noise of the measurements, although the qualitative time evolution is still well reproduced. It seems that the measurement error mainly acts as a blurring of the heatmaps. However, when considering the normalized correlators $\Gamma_{ij} / \max_{ij} \Gamma_{ij}$ in Figure~\ref{fig:PostselectCorrelations} and comparing them to the post-selected results in Figure~\ref{fig:correlations}, we notice a strong qualitative difference. This is consistent with the observable $N_{\text{flip}}$ without post-selection in Figure~\ref{fig:PostselectNFflip}(a) failing to reproduce the qualitative behavior of the model.

In Figure~\ref{fig:PostselectNFflip}(b) we show the fraction of discarded shots
\begin{equation}
    R_{\text{post}} = \frac{\text{\# discarded shots}}{\text{\# total shots}} \ .
\end{equation}
We find that roughly 50-70 \% of the shots are dropped by the post-selection for most time steps and values of $\delta$. The percentages seem to be stable up to a slight increase over simulation time $t$. This ratio is high, but still leaves $\sim 90-150$ shots per time step to compute expectation values.

\section{Full numerical model pipeline}
\label{sec:AppendixPostselection}
From the device, we obtain the measurement data in the form of strings that indicate if an atom was detected to be in the ground or the Rydberg state. Note that the Hamiltonians we aim to simulate preserve the particle number, i.e. the number of domain walls. We can easily count the number of domain walls in each shot, which allows us to apply a mild form of post-selection by discarding all shots with an incorrect number of domain walls. The data shown in the main text are computed from the post-selected data. A comparison of the raw data with the post-selected data can be found in Appendix \ref{sec:AppendixB}. Here, we focus on the density matrices that numerically represent our error model, including the post-selection.\\
The post-selection process can be modeled as a projector $P$ onto the subspace with the appropriate number of excitations / domain walls with subsequent renormalization of the state, which is applied after the time evolution.

Let $\rho_0 = \ket{g}^{\otimes {N-1}} \bra{g}^{\otimes N-1}$ be the pure initial state of the simulator. We denote with $U_{Ryd}$ the time evolution under the Rydberg Hamiltonian with ideal atom positions and ideal laser when executing the pulse sequence shown in Figure~\ref{fig:register}(d). $U_{Ryd, m}$ denotes the same time evolution. However, the subscript $m$ indicates that this is the $m$-th disorder realization. Then we can define the time evolved states for the ideal Rydberg or ideal Ising simulator
\begin{equation}
    \rho_{\text{ideal}} \sim P U_{Ryd} \rho_0 U_{Ryd}^\dagger P \ ,
\end{equation}
while the Rydberg simulator state with disorder but without RO error is given by
\begin{equation}
    \rho_{\text{no-RO}} \sim \frac{1}{M} \sum_{m=1}^M P U_{Ryd,m} \rho_0 U_{Ryd,m}^\dagger P \ ,
\end{equation}
where we perform $M$ unitary simulations and average the results. The full error model with positional disorder, laser imperfections and RO error is given by additionally applying the RO channel introduced in Appendix~\ref{sec:AppendixErrorModel}, i.e.
\begin{equation}
    \rho_{\text{error}} \sim \frac{1}{M} \sum_{m=1}^M P D_{RO} \left[ U_{Ryd,m} \rho_0 U_{Ryd,m}^\dagger \right] P \ .
\end{equation}
Note that due to the projection that models the post-selection, the density matrices after applying $P$ are not normalized, which would be required to obtain valid statistics. Therefore, after renormalization $\rho' = \rho / \Tr(\rho)$, the expectation value of the observable $A$ is then simply given by $\langle A \rangle = \Tr(A\rho')$.

\bibliography{references}

\end{document}